# Feshbach resonance and mesoscopic phase separation near a quantum critical point in multiband FeAs-based superconductors

Rocchina Caivano[1], Michela Fratini[1], Nicola Poccia[1], Alessandro Ricci[1], Alessandro Puri[1], Zhi-An Ren[2], Xiao-Li Dong[2], Jie Yang[2], Wei Lu[2], Zhong-Xian Zhao[2], Luisa Barba[3], Antonio Bianconi[1]

*1. Department of Physics, Sapienza University of Rome, P. Aldo Moro 2, 00185 Roma, Italy*

*2. Laboratory for Superconductivity, Institute of Physics and Beijing National Laboratory for Condensed Matter Physics, Chinese Academy of Sciences, P. O. Box 603, Beijing 100190, P. R. China*

*3. Institute of Crystallography, National Council of Research, Elettra, 34012 Trieste, Italy*

**Abstract**: High $T_c$ superconductivity in FeAs-based (pnictides) multilayers, evading temperature decoherence effects in a quantum condensate, is assigned to a Feshbach resonance (called also shape resonance) in the exchange-like interband pairing. The resonance is switched on by tuning the chemical potential at an electronic topological transition (ETT) near a band edge, where the Fermi surface topology of one of the subbands changes from 1D to 2D topology. We show that the tuning is realized by changing i) the misfit strain between the superconducting planes and the spacers ii) the charge density and iii) the disorder. The system is at the verge of a catastrophe i.e. near a structural and magnetic phase transition associated with the stripes (analogous to the 1/8 stripe phase in cuprates) order to disorder phase transition. Fine tuning of both the chemical potential and the disorder pushes the critical temperature $T_s$ of this phase transition to zero giving a quantum critical point. Here the quantum lattice and magnetic fluctuations promote the Feshbach resonance of the exchange-like anisotropic pairing. This superconducting phase that resists to the attacks of temperature is shown to be controlled by the interplay of the hopping energy between stripes and the quantum fluctuations. The superconducting gaps in the multiple Fermi surface spots reported by the recent ARPES experiment of D. V. Evtushinsky et al. arXiv:0809.4455 are shown to support the Feshbach scenario.

Key words: Multiband anisotropic superconductivity; interband pairing; Feshbach resonance, shape resonance; quantum coherence, temperature decoherence effects; FeAs superconductors.

## 1. Introduction

The discovery of high temperature superconductivity in FeAs multi-layered materials [1-10] provides a new system where the unknown quantum mechanism for evading temperature decoherence effects in a macroscopic quantum condensate is active. In fact in these FeAs based materials the superconducting quantum coherent phase of a fermionic gas resists at temperatures higher than the liquid hydrogen boiling temperature like in doped cuprate perovskites and diborides. It is possible that there is common quantum mechanism active in all these materials therefore the research is looking for similarities of the normal and superconducting phase between these systems. A common structural characteristic of cuprates, diborides, and FeAs-based superconductors is the heterostructure at the atomic limit: a superlattice of metallic layers with strong covalent bonds (atomic $CuO_2$ bcc layers or atomic graphene-like $B_2$ monolayers, or molecular $FeAs_4$ layers) intercalated by spacers made of different materials with a different electronic structure (fcc rocksalt layers $La_2O_2$ [11,12], or hcp Mg/Al layers, or rare earth oxide layers respectively) [13].

There are different types of layered FeAs-based superconductors that have an analogous structure.

a) Doped quaternary rare-earth iron oxypnictides, ROFePn (R = rare-earth metal and Pn = pnicogen = O) (R=La,Pr,Nd,Ce,Sm…) made of FeAs layers intercalated by RO oxide layers [1-10]. These "1111" systems at room temperature have a tetragonal (space group P4/nmm) structure. It is critical to the high $T_c$ superconductivity (55 K is the maximum $T_c$) the F substitution for oxygen (15-20 atm%, called electron doping (n-type) of the formal $[FeAs]^{-1}$ layers; or $Sr^{2+}$ for $R^{3+}$ doping (4-13 atm%, called hole doping (p-type); or the introduction of oxygen defects.

b) Doped alkaline earth iron arsenides, $AeFe_2As_2$ (Ae=Sr,Ba), made of $[Fe_2As_2]^{-2}$ layers separated by simple Ae-layers have a tetragonal $ThCr_2Si_2$-type, space group I4/mmm called "122". They become superconductors (38 K maximum $T_c$) with appropriate substitution of bivalent Ae cations with monovalent alkali metals, K, Cs, etc… For example the K for Sr substitution of 45 atm% in $Sr_{1-x}K_xFe_2As_2$ gives the maximum $T_c$ [14-19].

c) Undoped compounds, $KFe_2As_2$ and $CsFe_2As_2$, made of $[Fe_2As_2]^{-1}$ layers separated by monovalent ions are superconducting, with low $T_c$'s of 3.8 K and 2.6 K.

d) Undoped LiFeAs made of $[FeAs]^{-1}$ layers is a superconductor with $T_c$ = 18 K

[20].

e) Undoped non-superconducting AeFe$_2$As$_2$ (Ae=Ca,Sr,Ba) compounds, made of [Fe$_2$As$_2$]$^{-1}$ layers, become superconductors under pressure. [21-23].

### 2. The Internal Pressure: misfit strain

There is clear evidence that the high T$_c$ superconductivity in these multi-layer systems occurs by tuning the chemical potential at a particular point of the electronic structure. The chemical potential can be tuned by different methods: 1) by changing the charge density in the active layers via the control of the charge transfer, from or to the active layers; this is achieved by substitution of ions in the spacer layers, with others having a different ionic charge; 2) by changing the lattice parameters of the heterostructure at atomic limit by external pressure or internal chemical pressure.

The chemical pressure in these superlattices is due to the lattice mismatch, called misfit strain, between the alternated layers. The misfit strain could control the bond distance in the active superconducting layers and could induce a disorder, due to the formation of dislocations [24], lattice stripes, and incommensurate lattice modulations [25]. The misfit strain is usually defined as $\eta=(a_1-a_2)/a$ where $a_1$ and $a_2$ are the unit cell parameters of the ideal first and second layers respectively when they are well separated and $a=(a_1+a_2)/2$.

The first layers of the superlattice exhibit a compressive $\varepsilon_{1c}=(a_1-a)/a$ microstrain and the second layers a tensile $\varepsilon_{2t}=(a-a_2)/a$ microstrain. The average strain is zero in the superlattice $\varepsilon=\varepsilon_{1c}=\varepsilon_{2t}$ and the lattice parameter of the superlattice is close to $a=(a_1+a_2)/2$. Therefore the internal chemical pressure (misfit strain or mismatch chemical pressure) acting on the active layers in the superlattice can be obtained by measuring the lattice parameter of the superlattice "a" and the unrelaxed ideal lattice parameter of only one of the two layers, in fact $\eta=\varepsilon_{1c}+\varepsilon_{2t}=2\varepsilon$.

The chemical pressure is changed in cuprates [26-30] and in diborides [13] by chemical substitution of ions with different ionic radii in the spacer layers (in manganites $\eta=1-t$ where t is the tolerance factor). The chemical pressure acts as a complex anisotropic stress tensor that produces a compressive microstrain of the bcc $CuO_2$ layer in cuprates [26-30] and a tensile microstrain of the graphene-like B layer in magnesium diboride [13]. In cuprates by increasing the chemical pressure at constant doping, $\delta=1/8$, a structural phase transition LTO-LTT occurs at a critical compressive misfit strain 8% [27-30]. In the proximity of this structural phase transition a nanoscale phase separation [31] in presence of quenched disorder and a commensurate-incommensurate transition [32] for mobile dopants have been observed. Also the lattice structure of magnesium diborides is in the verge of a catastrophe [13]. The misfit strain of the FeAs quasi 2D layers can be easily measured in fact it is made of edge sharing $FeAs_4$ tetrahedral units where the FeAs bond length remains constant, $R_0=240$ pm, therefore the chemical mismatch pressure induces only a rotation of the bonds pushing the As-Fe-As bond out of the ideal value of the tetrahedral angle 109°28' [8] where the ideal lattice parameter of the orthorhombic lattice is $a_o=\sqrt{2}a_T =542.7$ pm. The misfit strain, measuring the chemical internal pressure, is therefore given by $\eta = 2(a_o/542.7 - 1) = 2(a_T/390.8 - 1)$.

In this work have measured the misfit strain of the undoped parent compounds of FeAs-based superconductors RFeAsO systems by powder X-ray diffraction. The ROFeAs (R=La, Pr, Nd and Sm) powder samples have been synthesized in Bejing as described elsewhere [3-5]. The X-ray diffraction patterns were recorded at the x-ray Diffraction beam-line at the Elettra synchrotron radiation facility in Trieste. The results of the measure of the misfit strain as a function of the ionic radius in the spacer layers are shown in Fig. 1 (panel a). The lattice parameters of the 122" systems shown in (panel b) are taken for the literature. It is clear from the data that the FeAs layers in all undoped "1111" and "122" systems suffer a tensile misfit strain. The chemical doping giving the superconducting phase doping not only changes the charge transfer from the spacer to the FeAs layers but also the misfit strain. The tensile misfit strain is reduced in "1111" and in "122" samples at optimum doping the misfit strain is close to zero [8]. A large tensile misfit strain promotes the low temperature charge and spin ordering phase that competes with superconductivity, and high $T_c$ superconductivity prevails where the misfit strain goes to zero. A compressive misfit strain shows up in low temperature superconductors made of "122" structure with intercalated monovalent alkali metal ions.

## 2. The Tetragonal to Orthorhombic Structural Transition

It is known that tuning the chemical potential at an electronic topological transition (ETT) the electron gas shows a 2.5 Lifshitz electronic topological transition; the compressibility of the electron gas becomes negative, therefore the system has the tendency toward a first order electronic phase separation; and finally the system can undergo a structural phase transition or a magnetic phase transition due to the freezing of a charge density wave (CDW) or a spin density wave (SDW) with the nesting vector connecting different portions of the Fermi surface. In this condition the system is near a magnetic, charge, orbital and lattice instability associated with the appearing of SDW, CDW, structural phase transitions and mesoscopic phase separation (MePhS). The electronic instabilities at the ETT's have been widely studied in the case of one-dimensional (1D) and two dimensional (2D) single band systems with the formation of 1D CDW and 2D CDW insulating phases respectively. A poorly studied case of interest here for the high $T_c$ cuprate superconductors is the onset of 1D CDW (and/or SDW) in a 2D electronic system, in fact the 1D CDW opens only partial gaps in the 2D Fermi surface, forming a striped electronic matter with pseudogaps, multiple quasi-1D subbands and 2D bands that below $T_c$ gives a multi-gap superconducting phase. The second complex case that has been object of very few investigations is the one relevant for $MgB_2$ and FeAs-based superconductors: the case of multiband systems, where the Fermi level is crossing several types of ETT's only in one of the bands.

All undoped parent compounds of the FeAs-based layered superconductors are multiband systems [33,42] where the chemical potential is self tuned to a particular point such that the system shows a lattice instability driven a Fermi surface nesting wavevector. In fact all undoped parent compounds show a similar tetragonal to orthorhombic transition occurring at low temperature $T_s$ detected by high resolution x-ray diffraction. We show in Fig. 2 the splitting of the "a" axis in stoichiometric ROFeAs "1111" systems at the structural transition from tetragonal (space group P4/nmm) to orthorhombic space group (Cmma) at low temperature, observed in the systems with R= La, Nd, Sm in agreement with previous works [43-50]. In Fig. 2 we report the XRD results for $AFe_2As_2$ "122" systems that show a similar structural transition from tetragonal $ThCr_2Si_2$-type, space group I4/mmm, to orthorhombic Fmmm space group [51-55]. In the orthorhombic phase a static stripe magnetic phase has been found. The structural transition takes place in a range of about 2 K in "122" systems, and it has been interpreted as first order transition [55] since it shows hysteretic behaviour, but it has been also identified as a second order transition from the investigation of the scaling of the order parameter [52]. The structural transition in the "1111" systems shows a continuous character over a large temperature range above and below the critical temperature [45,46]. This structural transition is therefore similar to the martensitic transition in alloys [56] and superconducting A15 compounds [57].

There is a strong coupling between magnetic and structural order parameters [58,59]. The spin ordering below the critical temperature Ts is driven by the low temperature orthorhombic phase and it shows a striped phase with the antiferromagnetic coupling in the direction of the long Fe-Fe bond (the orthorhombic $a_o$ axis) and ferromagnetic coupling in the direction of the short Fe-Fe bond (the orthorhombic $b_o$ axis). Therefore this striped magnetic phase shows that the ordering of spin is coupled with the ordering of a lattice distortion (Fe-Fe short and long bonds) that is similar to the striped phase at 1/8 doping in cuprates associated with ordering of long and short Cu-O bonds [30]. The results in Fig. 2 clearly show that the critical temperature $T_c$ of the structural phase transition decreases with decreasing the tensile strain due to lattice misfit. It is possible to see that the $BaFe_2As_2$ case shows an anomalous behaviour.

**3. The mesoscopic phase separation at the orthorhombic to tetragonal structural transition.**

The superconducting phase is observed to emerge from the non-superconducting magnetically ordered phase through appropriate doping of the charge reservoir spacer blocks. Due to the complex Fermi surface of the FeAs-superconductors [41,42] the effects of doping on the electronic and superconducting properties are not clear. The FeAs-based materials are quite different from cuprates since the parent compounds are metallic systems and not Mott insulators. There is on the contrary a strong analogy with the high-$T_c$ cuprates if one assumes that the parent compound of all cuprates superconductors is the striped phase, at 1/8 doping and 8% misfit strain. In fact a few number of authors [30,60-66] have proposed that the relevant quantum critical point for high $T_c$ superconductivity in cuprates is where the superconducting phase competes with the striped phase, at 1/8 doping and 8% misfit strain.

Band structure calculations reveal that the Fermi surface includes electron as well as hole pockets. For superconducting $K_{1-x}Sr_xFe_2As_2$ near the optimal chemical substitution ($x_{optimum}$) the Hall coefficient was found to be positive hinting that the majority carriers are holes [67]. However, with complete substitution of x=1, the negative Hall coefficient of Sr-122 indicates that electrons dominate the transport properties [68] therefore increasing x introduces more electrons into the $Fe_2As_2$ layer and the chemical potential crosses the ETT's of these multiband system. These results are consistent with recent measurements [68,69] and band structure calculations [34]. We show in Fig. 3 the variation of $T_c$ and $T_s$ for $K_{1-x}Sr_xFe_2As_2$ from ref. 70. Furthermore, the pressure-induced superconductivity in the non-superconducting compounds $AeFe_2As_2$ (Ae=Ca,Sr,Ba) indicates the role of the lattice in tuning the chemical potential. Therefore the high $T_c$ phase can be reached by varying the lattice parameters (modified by the external pressure or internal pressure) and the carrier densities in the $Fe_2As_2$ layers. The pressure experiments in $K_{1-x}Ba_xFe_2As_2$ [23] and $K_{1-x}Sr_xFe_2As_2$ [70] show that the critical temperature is a function of both lattice parameters and charge density in the active FeAs layers and the maximum $T_c$ occurs along a line of points of charge density and pressure.

There is now a large agreement that by using pressure, internal pressure (as it is shown in Fig. 2) and doping (as it is shown in Fig. 3) it is possible to decrease the temperature $T_s$ of the structural and magnetic phase transition toward zero. For the case $K_{1-x}Sr_xFe_2As_2$ the system reaches a quantum critical point, where $T_s=0$ K, at a critical internal pressure (misfit strain), a critical charge density in the Fe 3d bands and a critical disorder. The superconducting critical temperature reaches a maximum x=0.55 as it is shown in Fig. 3.

The system shows a mesoscopic phase separation (MePhS) of orthorhombic striped magnetic clusters and tetragonal superconducting clusters in the proximity of the quantum critical point for the structural phase transition in Fig. 3. We show a pictorial view of this MePhS in Fig. 4 in a first high doping regime, where the average structure is the tetragonal lattice, and in a second low doping regime, where the average structure is the orthorhombic lattice. In the orthorhombic clusters the charges can move only along the stripes in the b direction and are localized by the magnetic interaction in the direction. Therefore the first superconducting regime can be called a case of nematic electronic phase of itinerant fluctuating striped bubbles.

Therefore in the proximity of the zero temperature transition from the average orthorhombic phase to the tetragonal phase (a quantum phase transition driven by charge density, chemical pressure or pressure) there should be a Fermi surface that fluctuates in space and time between a 2D topology in the tetragonal clusters and a 1D topology in the orthorhombic clusters.

**3. The Feshbach resonance in a fluctuating striped phase**

We propose for FeAs-based superconductors a pairing mechanism, called the "shape resonance" or "Feshbach resonance" scenario that has been proposed for the cuprates [64-66] and diborides [71] and it is similar to the pairing mechanism in ultracold gases called "Feshbach resonance". The key interband pairing process is a Kondo exchange-like interaction between *first pairs* (with spin up and spin down) in a first Fermi surface portion and the *second pairs* in a second Fermi surface portion and are distributed in different spatial locations.

The "shape resonances" have been described by Feshbach in elastic scattering cross-section for the processes of neutron capture and nuclear fission [72] in the cloudy crystal ball model of nuclear reactions. This scattering theory is dealing with configuration interaction in multi-channel processes involving states with *different spatial locations.* Therefore these resonances can be called Feshbach shape resonances. These resonances are a clear well established manifestation of the non-locality of quantum mechanics and appear in many fields of atomic physics [73] and chemistry such as the molecular association and dissociation processes [74,75]. Feshbach resonances for molecular association and dissociation have been proposed for the manipulation of the interatomic interaction in ultracold atomic gases. In fact the interparticle interaction shows resonances tuning the chemical potential of the atomic gas around the energy of a discrete level of a biatomic molecule controlled by an external magnetic field [76]. This quantum phenomenon has been used to achieve the Bose-Einstein condensation (BEC) in the dilute bosonic gases of alkali atoms [77] and to get a BCS-like condensate in fermionic ultra-cold gases with large values of $T_c/T_F$ [78]. The process for increasing $T_c$ by a Feshbach resonance was first proposed by Blatt and Thompson [79-81] in 1963 for a superconducting thin film and it was called by Blatt "shape resonance". Blatt described the shape resonance in a superconducting thin film of thickness L where the chemical potential crosses the bottom $E_n$ of the n-th subband of the film, a quantum well, characterized by $k_z = n\pi/L$ with n>1. Therefore it occurs where the chemical potential $E_F$ is tuned near the critical energy $E_F=E_n$ for a 2.5 Lifshitz electronic topological transition (ETT) [82] at a band edge. At this ETT a small Fermi surface of a second subband disappears while the large 2D Fermi surface of a first subband shows minor variations. In the "clean limit" the single electrons cannot be scattered from the n-th to the (n-1)-th subband and viceversa because of disparity and negligible spatial overlap but configuration interaction between pairs in different subbands is possible in an energy window around $E_F=E_n$. Therefore the Feshbach shape resonance occurs by tuning the Lifshitz parameter $z=E_F-E_n$ around $z=0$. In the Blatt proposal z is tuned by changing the film thickness. The prediction of Blatt and Thompson of the oscillatory behavior of $T_c$ as a function of film thickness L has been recently confirmed experimentally for a superconducting film [83] although phase fluctuations due to the electron confinement in the two dimension is expected to reduce the critical temperature.

We have proposed to increase $T_c$ via a Feshbach or shape resonance not in a single layer but in a multilayer (or superlattice) made of superconducting layers intercalated by spacer layers [84-93] in the proximity of a quantum critical point. This proposal was advanced following the experimental evidence that in cuprates made of a superlattices of $CuO_2$ layers intercalated by rocksalt spacer layers, the $CuO_2$ plane shows nanoscale striped lattice fluctuations detected by EXAFS with a time scale of $10^{-15}$ sec [84], that is a signature for the proximity to the structural critical point for the LTO to LTT phase transition.

To describe the basic physics of superconductivity in these systems we have to overcome the approximations of the standard BCS theory for homogeneous systems considering an effective single band and focus on multiband superconductivity and anisotropic pairing mechanisms. In fact the standard BCS approximation assumes the Fermi energy at an infinite distance from the top or the bottom of the conduction band, and the pairing mechanism is not electronic state dependent (*the isotropic approximation*). The BCS wave-function of the superconducting ground state is constructed by configuration interaction of all electron pairs (+k with spin up, and -k with spin down) on the Fermi surface in an energy window that is the energy cut off of the interaction,

$$|\Psi_{BCS}\rangle = \prod_k (u_k + v_k c^+_{k\uparrow} c^+_{-k\downarrow})|0\rangle \qquad (1)$$

where $|0\rangle$ is the vacuum state, and $c^+_{k\uparrow}$ is the creation operator for an electron with momentum k and spin up.

In anisotropic superconductivity one has to consider configuration interaction between pairs, in an energy window $\Delta E$ around the Fermi level, in different locations of the k-space with a different pairing strength, that gives a k-space dependent superfluid order parameter i.e., a k-dependent superconducting gap. A particular case of anisotropic superconductivity is multiband superconductivity, where the order parameter and the excitation gap are mainly different in different bands. The theory of two band superconductivity, including the configuration interaction of pairs of opposite spin and momentum in the *a*-band and *b*-band the many body wave function is given by

$$|\Psi_{Kondo}\rangle = \prod_k (u_k + v_k a^+_{k\uparrow} a^+_{-k\downarrow}) \prod_k (x_k + y_k b^+_{k\uparrow} b^+_{-k\downarrow})|0\rangle \qquad (2)$$

The element corresponding to the transfer of a pair from the *a*-band to the *b*-band or vice versa appears with the negative sign in the expression of the energy. This gain of energy is the origin of the increase of the transition temperature driven by interband pairing. The two-band superconductivity has been proposed for metallic elements and alloys [94-119], for doped cuprate perovskites [120-152], for magnesium diboride [71,153-178] and for few other materials as Nb doped $SrTiO_3$ [179], $Sr_2RuO_4$ [180-181] $YNi_2B_2C$, $LuNi_2B_2C$ [182] and $NbSe_2$ [183] and superlattices of carbon nanotubes [184].

The multiband superconductivity shows up only in the "clean limit", where the single electron mean free path for the interband impurity scattering satisfies the condition $l > \hbar v_F / \Delta_{av}$ where $v_F$ is the Fermi velocity and $\Delta_{av}$ is the average superconducting gap. Therefore the criterion that the mean free path should be larger than the superconducting coherence length must be met. This is a very strict condition that implies also that the impurity interband scattering rate $\gamma_{ab}$ should be very small $\gamma_{ab} \ll (1/2)(K_B/\hbar)T_c$. Therefore most of the metals are in the "dirty limit" where the interband impurity scattering mixes the electron wave functions of electrons on different spots on the bare Fermi surfaces and it reduces the system to an effective single Fermi surface.

The "interchannel pairing" or "interband pairing" that transfers a pair from the "a"-band to the "b"-band and vice versa in the multiband superconductivity theory is expressed by the off diagonal element

$$\sum_{k,k'} J(k,k')(a^+_{k\uparrow}a^+_{-k\downarrow}b_{-k\downarrow}b_{k\uparrow}) \qquad (3)$$

where $a^+$ and $b^+$ are creation operators of electrons in the "a" and "b" band respectively and $J(k,k')$ is an exchange-like integral. This interband pairing interaction may be repulsive as it was first noticed by Kondo [97]. Therefore it is a non-BCS pairing process since in the BCS theory an attractive interaction is required for the formation of Cooper pairs. Another characteristic feature of multiband superconductivity is that the order parameter shows the sign reversal in the case of a repulsive interband pairing interaction.

The non-BCS nature of the interband pairing process is indicated also by the fact that, when it is dominant, the isotope effect vanishes even if the intra-band attractive interaction in each band is due to the electron-phonon coupling. Moreover the effective repulsive Coulomb pseudopotential in the Migdal-Eliashberg theory is expected to decrease (so that the effective coupling strength increases) where the interband pairing is dominant.

## 5. The Feshbach resonance at the topological electronic transitions in a simple multiband system

FeAs-based superconductors, like all other superconductors, are multiband systems where the Fermi level in one of the bands is close to a band edge, a hole-electron van Hove singularity, or a 2D to 3D electronic topological transition or a 1D to 2D electronic topological transition. Therefore here there is a breakdown of both "the infinite Fermi energy approximation" and "the single band approximation" for the standard BCS for low temperature superconductors. This is a common feature for all high $T_c$ superconductors known so far. Therefore the superconducting order parameter is expected to be dependent on the details of the electronic structure and k-dependent as it is observed in these anisotropic multiband systems. The anisotropic multiband scenario introduces new possible terms in the pairing process where the superconducting condensate is formed taking advantage of the coulomb interactions between the fermions itself, via exchange of spin fluctuations between nested portions of the Fermi surfaces or acustic plasmons between different portions of the Fermi surfaces. The Feshbach resonance due to exchange of pairs between different portions of the Fermi surfaces could be the key mechanism for making a quantum condensate that avoids temperature de-coherence effects. In fact in a multiband superconductors there are some special conditions where the exchange-like interband pairing could show the Feshbach resonance.

Here we discuss a particular case of multiband superconductivity that grabs some key feature of Feshbach resonances in FeAs-based superconductors. We consider a toy electronic structure model: the case of a superlattice of quantum wires that simulates the electronic structure of a striped metallic and magnetic phase. Here the charge carriers in the superconducting layer move as free charges in the x direction (the short $b_o$-axis in the FeAs$_{4/4}$ 2D lattice) but they have to overcome a periodic potential barrier V(x,y), with period $\lambda_p$, amplitude $V_b$ and width W along the y direction (the long $a_o$-axis in the FeAs$_{4/4}$ 2D lattice) constant in the x direction, expressed for x=constant as:

$$V(y) = -V_b \theta\left(\frac{L}{2} - \tilde{y}\right) \quad \text{where} \quad \tilde{y} = y - q\lambda_p - \frac{\lambda_p}{2} \tag{4}$$

and q is the integer part of $y/\lambda_p$.

The solution of the Schrödinger equation for this system,

$$-\frac{\hbar^2}{2m}\nabla^2 \psi(x,y) + V(x,y)\psi(x,y) = E\psi(x,y)$$

where

$$\psi_{n,k_x,k_y}(x,y) = e^{ik_x x} \cdot e^{ik_y q\lambda_p} \psi_{n,k_y}(y) \tag{5}$$

in the stripe is given by

$$\psi_{n,k_y}(y) = \alpha e^{ik_w \tilde{y}} + \beta e^{-ik_w \tilde{y}} \quad for\ |\tilde{y}| < L/2$$

$$k_w = \sqrt{2m_w(E_n(k_y) + V_b)/\hbar^2}$$

in the barrier us given by

$$\psi_{n,k_y}(y) = \gamma e^{ik_b \tilde{y}} + \delta e^{-ik_b \tilde{y}} \quad for\ |\tilde{y}| \geq L/2$$

$$k_b = \sqrt{2m_b E_n(k_y)/\hbar^2}$$

The coefficients α, β, γ and δ are obtained by imposing the Bloch conditions with periodicity $\lambda_p$, the continuity conditions of the wave function and its derivative at L/2, and finally by normalization in the surface unit. The solution of the eigenvalue equation for E gives the electronic energy dispersion for the n subbands with energy $\varepsilon_n(k_x, k_y) = \varepsilon(k_x) + E_n(k_y)$ where $\varepsilon(k_x) = (\hbar^2/2m) k_x^2$ is the free electron energy dispersion in the x direction and $E_n(k_y)$ is the dispersion in the y direction.

There are $N_b$ solutions for $E_n(k_y)$, with $1 \leq n \leq N_b$, for each $k_y$ in the Brillouin zone of the superlattice giving a dispersion in the y direction of the $N_b$ subbands with $k_x=0$.

By changing the charge density it is possible to cross the ETT where the Fermi surface topology of the second subband changes form 2D to 1D as shown in Fig. 5. The partial density of states (DOS) of the n-th subband gives a step-like increase of the total DOS when the chemical potential reaches the bottom of the subband n=2 at energy $E_0$ where an ETT (appearing of a new Fermi surface spot) occurs as it is shown in Fig. 6. At the change from 2D to 1D topology a peak in the DOS is observed as it is shown in Fig. 6.

The superlattice induces a relevant <u>k dependent</u> interband pairing interaction $V_{n,n'}(k,k')$ that is the exchange-like non BCS interband pairing interaction. The interband interaction is controlled by the details of the *quantum superposition of* *states* between the wave functions of the pairing electrons in the different subbands of the superlattice

$$V_{n,n'}(k,k') = V^o_{n,k_y;n',k'_y}\theta(\hbar\omega_0 - |\varepsilon_n(k) - \mu|)\,\theta(\hbar\omega_0 - |\varepsilon_{n'}(k') - \mu|) \tag{6}$$

where $k = (k_x, k_y)$ and

$$V^o_{n,k_y;n',k'_y} = -J\int_S dx\,dy\,\psi_{n,-k}(x,y)\psi_{n',-k'}(x,y)\psi_{n,k}(x,y)\psi_{n',k'}(x,y)$$

$$= -J\int_S dx\,dy\,|\psi_{n,k}(x,y)|^2\,|\psi_{n',k'}(x,y)|^2$$

where n and n' are the subband indexes. $k_x$ ($k_x'$) is the component of the wavevector in the wire direction (or longitudinal direction) and $k_y$ ($k_y'$) is the superlattice wavevector (in the transverse direction) of the initial (final) state in the pairing process, and μ is the chemical potential.

In the separable kernel approximation, the gap parameter has the same energy cut off $\hbar\omega_o$ as the interaction. Therefore it takes the values $\Delta_n(k_y)$ around the Fermi surface in a range $\hbar\omega_o$ depending from the subband index and the superlattice wavevector $k_y$.

The self consistent equation, for the ground state energy gap $\Delta_n(k_y)$ is:

$$\Delta_n(\mu, k_y) = -\frac{1}{2N}\sum_{n',k'_y,k'_x}\frac{V_{n,n'}(k,k')\cdot\Delta_{n'}(k'_y)}{\sqrt{(E_{n'}(k'_y) + \varepsilon_{k'_x} - \mu)^2 + \Delta^2_{n'}(k'_y)}} \tag{7}$$

where N is the total number of wavevectors. Solving iteratively this equation gives the anisotropic gaps dependent on the subband index and weakly dependent on the superlattice wavevector $k_y$. The structure in the interaction gives different values for the gaps $\Delta_n$ giving a system with an anisotropic gaps in the different segments of the Fermi surface.

The superconducting gaps in the second, $\Delta_2$, and first, $\Delta_1$, subband in a superlattice of quantum wires are shown in Fig 5.

The increase of the gap $\Delta_1$ is driven only by the Feshbach resonance in the interband pairing since the partial DOS of the first subband has not peaks.

The critical temperature $T_c$ of the superconducting transition can be calculated by iterative method

$$\Delta_n(k) = -\frac{1}{N}\sum_{n'k'} V_{nn'}(k,k') \frac{\tgh(\frac{\xi_{n'}(k')}{2T_c})}{2\xi_{n'}(k')} \Delta_{n'}(k') \qquad (8)$$

where $\xi_n(k) = \varepsilon_n(k) - \mu$.

The interband pairing term enhances $T_c$ by tuning the chemical potential in an energy window around the Van Hove singularities, "z"=0, associated with a change of the topology of the Fermi surface from 1D to 2D of the second subband of the superlattice.

The critical temperature $T_c$ and the superconducting gap in the first 1D subband and in the second 2D subband are plotted in Fig. 6.

The chemical potential is normalized to the cut off energy $\omega_0$ that is the energy window around the Fermi energy where are the electrons pairs that contribute to the formation of the macroscopic quantum condensate.

In the FeAs-based superconductors near the quantum critical point for the Lifshitz ETT transition, the Van Hove feature in the electronic energy spectrum associated with a change of Fermi surface topology from 1D to 2D fluctuates around the Fermi level due to quantum fluctuations. The amplitude of the energy fluctuations controls the energy window where are located the electron pairs that contribute to the formation of the macroscopic quantum condensate wavefuction. Therefore the cut-off energy $\omega_o$ for the pairing is related with the quantum fluctuations.

We think that our calculations in Fig. 6 reproduce the basic experimental results of Fig. 3 and provide a qualitative understanding of the multigap superconductivity in FeAs superconductors. In the low temperature orthorhombic magnetic striped phase the electronic structure should be similar to the one described here where the Fermi level is in the pseudogap where all bands have a quasi 1D character. In this regime the static magnetism prevails on the superconducting order. Doping the system changes the position of the chemical potential and increases the randomness of the system up to a critical point where the first order phase transition becomes a continuous first transition with a fluctuating nanoscale phase separation like it is shown in Fig. 3. The superconducting phase emerges where at least one subband reaches the 1D to 2D electronic topological transition. Considering the Fermi surface of superconducting, $T_c$=32K, $Ba_{1-x}K_xFe_2As_2$ reported recently [42] we identify this subband with the *"inner $\Gamma$ barel"*. The *"inner $\Gamma$ barel"* is simulated by the *second subband* in our toy theoretical model described above. In fact the *"inner $\Gamma$ barel*" is clearly a small 2D Fermi close to the band edge and it shows a large superconducting gap of 9 meV and $2\Delta/K_BT_c = 6.8$. The *"outer $\Gamma$ barel*" be simulated by the first subband where the Fermi level is far from the band edge and the superconducting gap is smaller than 3 meVand $2\Delta/K_BT_c$ <3.

The Fermi surface spots called *"inner $\Gamma$ barel"* and the *"outer $\Gamma$ barel"* should be strongly modulated by quantum fluctuations being the system close to the zero temperature phase transition from the orthorhombic to tetragonal structural transition and from the static striped phase to the superconducting phase shown in Fig.3 with dynamical lattice and spin fluctuations as shown in Fig. 4.

Therefore we expect that in our toy model, shown ion Fig. 6, the onset of superconductivity shows up by moving the chemical potential across the 1D to 2D electronic topological transition. The superconducting critical temperature is different from zero where the chemical potential is in an energy range window around the ETT that in our model is the energy cut off. In this regime the superconducting condensate is made of configuration interactions including both electrons pairs in a quasi 1D Fermi surface and pairs in the 2D Fermi surface of the second subband as it is shown in Fig. 5.

Finally the chemical potential crosses the bottom of the band and the systems goes in the Bose like regime where all electrons in the second subband form the condensate. Therefore in this scenario where the Fermi level is tuned from the 1D to the 2D topology to it will be the regime of BCS-Bose crossover that is typical of cuprates and FeAs superconductors that follows the Uemura plot.

Finally in Fig. 7 we report the variation of the ratio $2\Delta/K_B T_{cmax}$ for the two gaps where the superconducting $T_c$ is maximum. The experimental values of the ratio $2\Delta/K_B T_{cmax}$ measured so far in FeAs superconductors is 6.8 and smaller than 3 showing a large deviation from the standard BCS value 3.5. In our calculations we show that the larger deviation from the BCS value occurs where $\omega_0 < D$.

In conclusion the new data discussed in this work point toward the Feshbach resonance of the exchange-like pairing at the ETT associated with the Fermi surface topology crossover as a possible scenario that grabs some key physics of the high superconductivity at the BCS to Bose crossover near a quantum critical point that is common between multiband cuprates [185], FeAs pnictides, diborides, and doped nanotubes [184]. We have provided evidence that the tuning of the chemical potential to the quantum critical point of the order-disorder transition of the stripes phase (analogous to the 1/8 stripe phase in cuprates) is controlled also by misfit strain beyond doping and disorder. The tuning drives the system at the verge of a catastrophe i.e. near the stripes order-to-disorder phase transition having a critical temperature $T_s$ going to zero. At this quantum critical point the quantum lattice and magnetic fluctuations promote the Feshbach resonance. The superconducting gaps are controlled by the interplay of the hopping between stripes and the quantum fluctuations. The recent detection of the multigaps by the ARPES experiment of Borisenko's group on the Fermi surface spots [42] are shown to support the Feshbach resonance scenario.

**Acknowledgments:** We thank the staff of the XRD beam line of Elettra synchrotron radiation facility in Trieste and Naurang L. Saini for help and discussions. We acknowledge financial support from European STREP project 517039 "Controlling Mesoscopic Phase Separation" (COMEPHS) (2005).

R. Caivano et al. arXiv:0809.4865 (2008)                      25

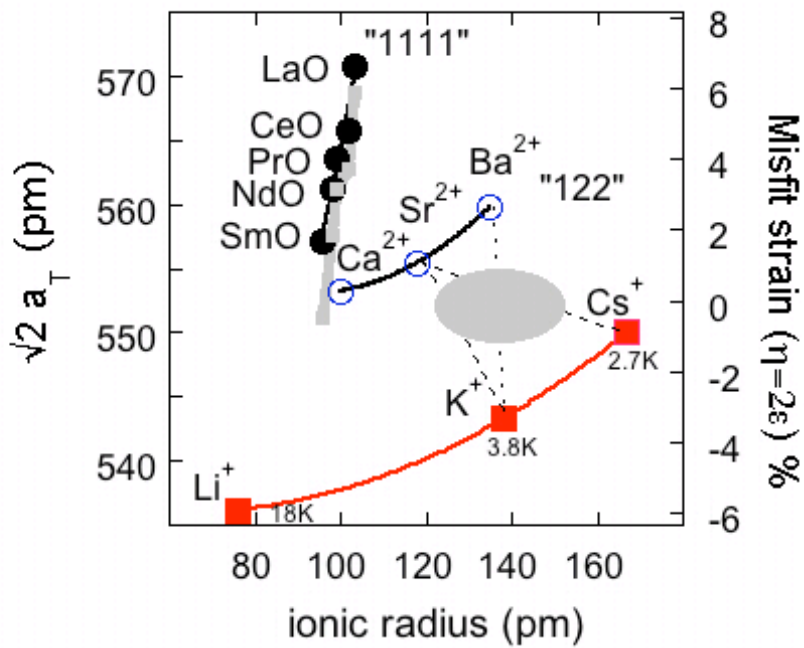

*Figure:1*. The values of the lattice parameter $a_O = \sqrt{2}a_T$, where $a_T$ is the "a" axis of the tetrahedral high temperature structure) of the FeAs-based stoichiometric parent compounds with the "1111" structure (filled circles) as a function of the ionic radius of the ions in the spacer layers measured by powder x-ray diffraction. The "$a_o$ axis" measuring two times the Fe-Fe distance for the system with the "122" structure having divalent alkaline earth ions (open circles) or monovalent alkali ions (filled squares) taken from the literature are reported, The chemical internal pressure or misfit strain on the left axis is given by $\eta = 2(a_O/542.17 - 1)$. The superconducting samples at optimum doping have a microstrain close to zero [8]. The high $T_c$ superconductivity shows up in doped systems located in the gray regions.

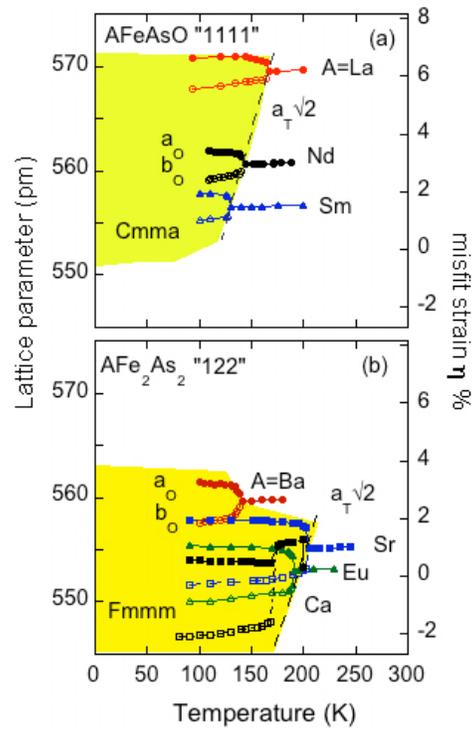

*Figure:2.* The structural parameter $a_O$, $b_O$ of the orthorhombic structure and $\sqrt{2}a_T$ of the tetragonal structure for the stoichiometric undoped parent compounds of the FeAs-based superconductors as a function of temperature, showing the structural phase transition from the high temperature tetragonal phase to the low temperature orthorhombic phase.

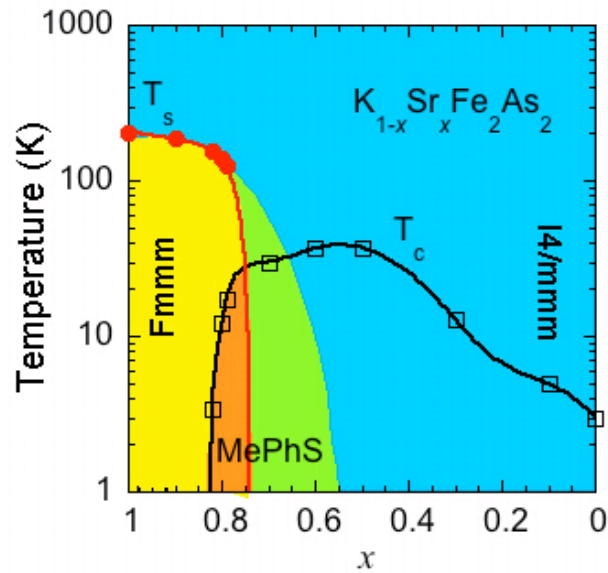

*Figure :3.* The superconducting critical temperature $T_c$ (open squares) for $K_{1-x}Sr_xFe_2As_2$ [70] as a function Sr content x. The critical temperature of the structural phase transition $T_s$ (filled circles) decreases to zero by decreasing x reaching a quantum critical point where the superconducting order in the tetragonal lattice competes with the magnetic striped phase in the orthorhombic lattice. The superconducting critical temperature reaches a maximum at x=0.55 and then decreases. A mesoscopic phase separation (MePhS) is clearly bserved in the tange 0.8<x<0.5 with coexisting tetragonal and orthorhombic nanoclusters.

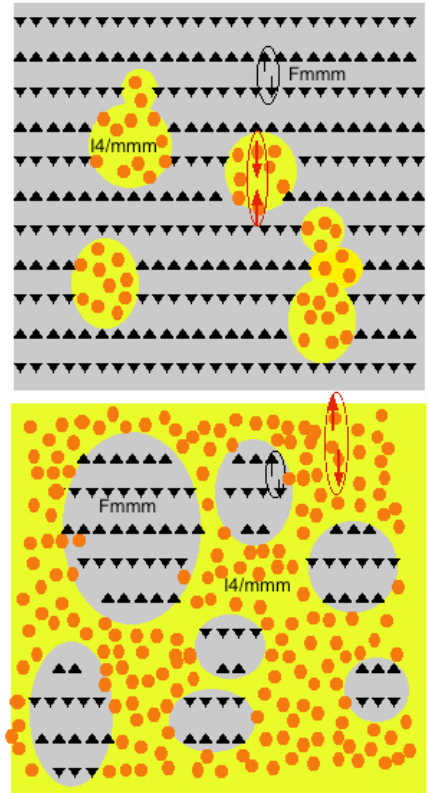

*Figure: 4*. Pictorial view of the fluctuating mesoscopic phase separation regime (MePhS) a) (upper panel) in the orhorhombic phase in the proximity of the structural phase transition from the orhorhombic (Fmmm) to the tetragonal (I4/mmm) structure where fluctuation nanoscale bubbles of metallic phase with a 2D Fermi surface (filled circles) coexists with the matrix of striped magnetic matter (triangles) with quasi 1D Fermi surface; and the scenario b) (lower panel in the tetragonal phase in the proximity of the structural phase transition from tetragonal (I4/mmm) to orhorhombic (Fmmm) structure where fluctuation nanoscale bubbles of striped magnetic matter (triangles) with quasi 1D Fermi surface coexists with the matrix of metallic phase (filled circles) with a 2D Fermi surface. The Feshbach resonance is described as the exchange of a pair of electrons in the 2D Fermi surface in the tetragonal phase with a pair of electrons in the striped matter with a quasi 1D Fermi surface.

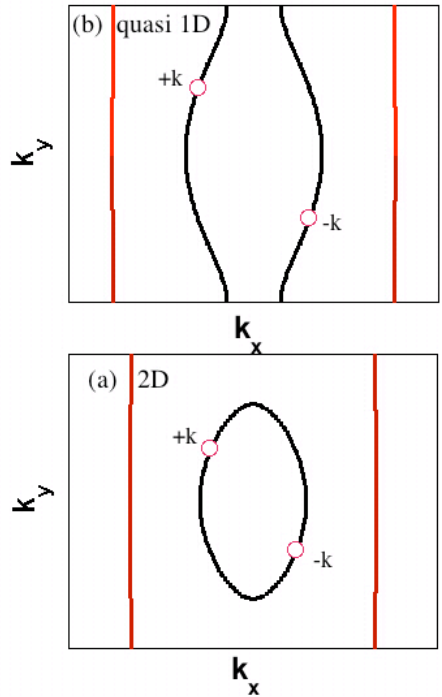

*Figure :5*. a) The Fermi surface above the bottom of the second subband made of a first 1D subband (vertical red lines) and a second 2D dimensional subband (black circle). By changing the pressure, charge density or misfit strain it is possible to cross a ETT where the DOS shows a sharp peak (see Fig. 5) and the second subband changes its topology from a 2D topology (panel a) to a 1D topology (panel b). A different type of ETT (appearing of a new Fermi surface spot) appears where the second subband disappears since it is not crossing the Fermi level.

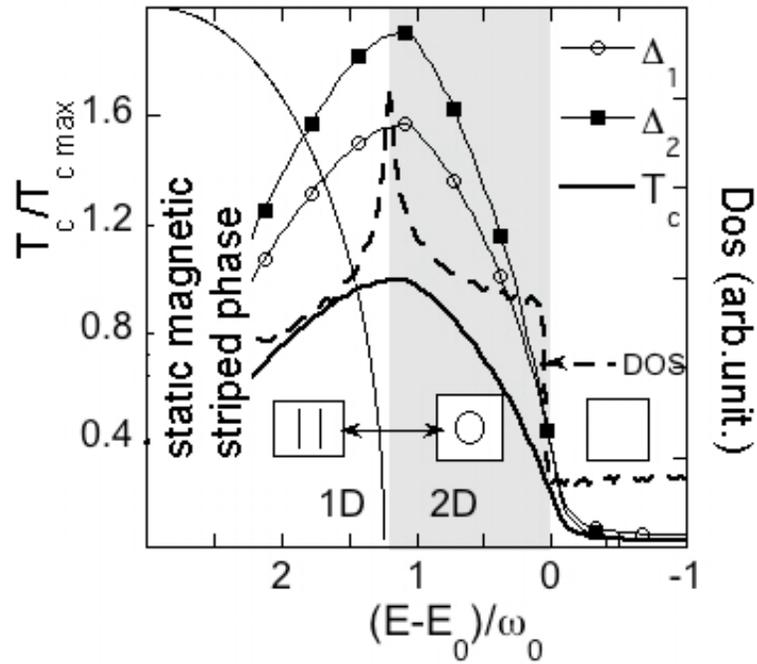

*Figure: 6.* The total density of states (DOS) for a superlattice of quantum wires near the bottom of the second subband (dashed line) for the case where the ratio between the pairing energy cut off ($\omega_0$) and the transversal energy dispersion (D) $\omega_0/D = 0.86$. The critical temperature $T_c$, the superconducting gaps in the first, $\Delta_1$, and second, $\Delta_2$, subband for a superlattice of quantum wires are normalized to the maximum $T_c$ and plotted as a function of the energy of the ratio of Fermi level minus the band edge and the energy cut off $\omega_0$.

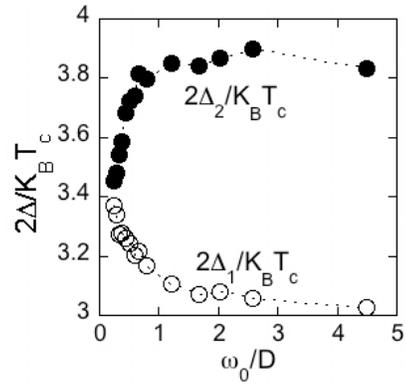

*Figure 7.* The ratio $2\Delta_1/K_BT_c$ and $2\Delta_2/K_BT_c$ as a function of the ratio between the pairing energy cut off ($\omega_o$) and the transversal energy dispersion (D) between stripes. The largest deviation of the ratio $2\Delta/K_BT_c$ from the expected BCS value 3.5 occurs for the large values of the ratio $\omega_o/D$ while for small $\omega_o/D<1$ it converges toward the BCS value.